\documentclass[twocolumn,aps,prb,amsmath,amssymb]{revtex4} %showpacs,
\usepackage{bm}
\usepackage{graphicx}
\newcommand{\iChEM}{{\it i}{\rm ChEM}}
\newcommand{\Sec}[1]{Sec.\,\ref{#1}}

\newcommand{\B}{\mbox{\tiny B}}
\newcommand{\tS}{\mbox{\tiny S}}
\newcommand{\tT}{\mbox{\tiny T}}
\newcommand{\D}{\mbox{\tiny D}}

\newcommand{\T}{\mbox{\tiny T}}
\newcommand{\SB}{\mbox{\tiny SB}}
\newcommand{\FP}{\mbox{\tiny FP}}

\newcommand{\w}{\omega}
\newcommand{\ti}{\tilde}
\newcommand{\nl}{\nonumber \\}

\newcommand{\wti}{\widetilde}

\newcommand{\be}{\begin{equation}}
\newcommand{\ee}{\end{equation}}
\newcommand{\bsube}{\begin{subequations}}
\newcommand{\esube}{\end{subequations}}
\newcommand{\Eq}[1]{Eq.\,(\ref{#1})}
\newcommand{\Eqs}[1]{Eqs.\,(\ref{#1})}

\newcommand{\dg}{\dagger}
\newcommand{\la}{\langle}
\newcommand{\ra}{\rangle}
\newcommand{\La}{\big\la}
\newcommand{\Ra}{\big\ra}

\begin{document}

\title{Theory of quantum dissipation in a class of non-Gaussian environments}

\author{Rui-Xue Xu}
\author{Yang Liu}
\author{Hou-Dao Zhang}
\author{YiJing Yan} \email{yanyj@ustc.edu.cn}

\affiliation{Hefei National Laboratory for Physical Sciences at the Microscale
and Department of Chemical Physics
and Synergetic Innovation Center of Quantum Information and Quantum Physics
and \iChEM,
%(Collaborative Innovation Center of Chemistry for Energy Materials)
University of Science and Technology of China, Hefei, Anhui 230026, China}

\date{June 22, 2017}

\begin{abstract}

  In this work we construct a novel dissipaton--equation--of--motion (DEOM)
theory in quadratic bath coupling environment,
based an extended algebraic statistical quasi--particle approach.
%[Y.~J.~Yan, J.\ Chem.\ Phys.\ {\bf 140}, 054105 (2014)].
%
To validate the new ingredient of the underlying dissipaton algebra, we derive
an extended Zusman equation via a totally different approach.
We prove that the new theory, if it starts with the
identical setup, constitutes the dynamical resolutions to the extended Zusman equation.
Thus, we verify %by \emph{de facto}
the generalized (non--Gaussian) Wick's theorem with dissipatons--pair added.
This new algebraic ingredient enables the dissipaton approach
being naturally extended to nonlinear coupling environments.
Moreover, it is noticed that, unlike the linear bath coupling case,
the influence of a non-Gaussian environment cannot be
completely characterized with the linear response theory.
The new theory has to take this fact into account.
The developed DEOM theory manifests the dynamical interplay between
dissipatons and nonlinear bath coupling descriptors that will be
specified.
Numerical demonstrations will be given with the optical line shapes
in quadratic coupling environment.

\end{abstract}

\maketitle

\section{Introduction}

  Quantum dissipation plays crucial roles in many fields of modern science.
Exact theories include the Feynman--Vernon influence functional
path integral approach,\cite{Fey63118}
and its differential equivalence,
the hierarchical--equations--of--motion (HEOM) formalism.
\cite{Tan906676,Tan06082001,Xu05041103,Yan04216,Jin08234703}
However, almost all existing quantum dissipation theories exploit the Gaussian--Wick's
thermodynamical statistics,\cite{Wei08,Kle09,Yan05187} which is strictly valid
only for linear bath couplings.
%%%
Intrinsically, a linear bath coupling implies
a \emph{weak backaction} of system on environment.
The lowest non--Gaussian environment influence
requires a quadratic bath coupling.

  In this work, we extend the equation of motion (DEOM)
theory,\cite{Yan14054105,Yan16110306}
to treat the linear--plus--quadratic bath coupling environment.
%%%%
This theory goes with a statistical quasi-particle (``dissipaton'')
description for the hybrid environment that can be either bosonic or fermionic
or excitonic.
Dynamical variables in DEOM are the dissipaton density operators (DDOs),
for both the reduced system and
and the hybrid bath dynamics.\cite{Yan14054105,Yan16110306}
The latter could also be
measured experimentally, via such as the Fano interference,
\cite{Fan611866,Mir102257,Zha15024112,Xu151816,Zha16237}
vibronic spectroscopy with non-Condon polarized environment,\cite{Zha16204109}
and transport current noise spectrum.\cite{Jin15234108}
%%%

  Dissipaton algebra plays essential roles here.\cite{Yan14054105,Yan16110306}
It consists of  the generalized (non--Gaussian) Wick's theorem
and the generalized diffusion equation.
%%%
This noval algebra leads to the rules on how the DDOs evolves in time,
and further on their relations to experimental measurable quantities
that involve explicitly the hybrid bath dynamics.%
\cite{Zha15024112,Xu151816,Zha16237,Zha16204109,Jin15234108}
%%%
From the algebraic construction point of view,
the new DEOM theory in quest for amounts to the establishment
of the generalized Wick's theorem with dissipatons--pairs added.
This will be the new ingredient of the dissipaton algebra
for treating the quadratic bath coupling in study.
%%%%

 Another important issue is concerned with the characterization of nonlinear coupling bath.
On top of the interacting bath
correlation function description,\cite{Wei08,Kle09,Yan05187}
addition information would be needed.
%%%
This is a general concern in any non--Gaussian environment theories.
%%%%
To address this issue, we adopt a polarization
model to determine both the linear
and nonlinear bath coupling strengths.\cite{Xu17JCP1}
This model resolves this issue,
with a single additional parameter, on top of the
conventional linear response theory.
%%%%

 This paper is organized as follows.
In \Sec{thsec2}, we construct the DEOM formalism,
via the dissipaton algebra, including the aforementioned new ingredient
for treating quadratic bath coupling.
%%%
In \Sec{thsec3}, we validate this new ingredient, the generalized
Wick's theorem with dissipatons--pairs added.
To do that we derive an extended Zusman  equation via a totally different approach.
We prove that the DEOM formalism, if it started with the
same setup, constitutes the dynamical resolutions to the extended Zusman equation.
%%%%
Therefore, as the algebraic construction is concerned,
we would have also confirmed the DEOM formulations presented in \Sec{thsec2}.
%%%%
Appeared there are also the linear and quadratic bath coupling strength
parameters, which will be discussed on the basis of the nonlinear polarization model
in \Sec{thsec4}.
%%%
Numerical DEOM demonstrations are then carried out
on the optical line shapes
in the nonlinear coupling environment.
We conclude in \Sec{thconc}.

\section{Dissipaton dynamics theory}
\label{thsec2}

\subsection{Statistical quasi--particle description}
\label{thsec2A}

 Let us start with the total composite Hamiltonian,
\be\label{HT}
  H_{\tT}=H_{\tS} + h_{\B}
  + \hat Q_{\tS}(\alpha_1\hat x_{\B}+\alpha_2\hat x_{\B}^2).
\ee
The system Hamiltonian $H_{\tS}$ and dissipative operator $\hat Q_{\tS}$
are arbitrary. The latter is set to be dimensionless.
The bath Hamiltonian and the hybridization bath operator (solvation mode)
are given, respectively,
\be\label{hB0}
   h_{\B}= \frac{1}{2}\sum_j \omega_j(\hat p_j^2 + \hat q_j^2)
\ \  \text{and} \ \
   \hat x_{\B}=\sum_j c_j \hat q_j,
\ee
 Throughout this paper we set $\hbar=1$ and $\beta=1/(k_{B}T)$,
with $k_B$ and $T$ being the Boltzman constant and temperature.
Let $\la \hat O \ra_{\B}\equiv {\rm tr}_{\B}\big(\hat O e^{-\beta h_{\B}}\big)/{\rm tr}_{\B}
e^{-\beta h_{\B}}$ be the bare bath ensemble average.
Define $\hat x_{\B}(t)\equiv e^{ih_{\B}t}\hat x_{\B} e^{-ih_{\B}t}$.
Set hereafter $t\geq 0$ for the time variable.
We have\cite{Yan05187,Wei08}
\begin{align}\label{FDT}
 \la \hat x_{\B}(t)\hat x_{\B}(0)\ra_{\B}
&= \frac{1}{\pi} \int^{\infty}_{-\infty}\! {\rm d}\omega
    \frac{e^{-i\omega t}\chi^{(i)}_{\B}(\w)}{1-e^{-\beta\omega}}
\nl&
  =\sum^K_{k=1} \eta_k e^{-\gamma_kt}.
\end{align}
The first expression is the fluctuation--dissipation theorem (FDT),\cite{Yan05187,Wei08}
with $\chi^{(i)}_{\B}(\w)$ being the imaginary part of
\be\label{chiw_def}
  \chi_{\B}(\w)\equiv i\int^{\infty}_{0}\!{\rm d}t\,e^{i\w t}
   \La [\hat x_{\B}(t),\hat x_{\B}(0)]\Ra_{\B}.
\ee
The second expression of \Eq{FDT} presents
an exponential series expansion of the
linear bath correlation function.
Set hereafter $\hat x_{\B}$ to be dimensionless, so that
the bath coupling parameters, $\alpha_1$ and $\alpha_2$ in \Eq{HT},
are of frequency unit.
 It is well known that, for a complete characterization of the nonlinear
environment influence ($\alpha_2\neq 0$),
additional information, on top of \Eq{FDT}, is needed.
We will address this issue in \Sec{thsec4}.

 Dissipatons, $\{\hat f_k\}$, arise strictly via the linear bath coupling part, as follows.
\be\label{xB_in_f}
 \hat x_{\B} = \sum_{k=1}^K \hat f_{k},
\ee
with
\be\label{ff_t}
\begin{split}
 \la \hat f_{k}(t)\hat f_{j}(0)\ra_{\B}&=\delta_{kj}\eta_{k}e^{-\gamma_k t}, \
\\
 \la \hat f_{j}(0)\hat f_{k}(t)\ra_{\B}&=\delta_{kj}\eta^{\ast}_{\bar k}e^{-\gamma_k t}.
\end{split}
\ee
The associated index $\bar k$ in the second expression above is defined via
$\gamma_{\bar k}\equiv \gamma^{\ast}_{k}$.
It is easy to verify that both \Eq{FDT} and
its time reversal are reproduced.
 Denote for later use
\be\label{ff_t0}
\begin{split}
  \la \hat f_{k}\hat f_{j}\ra^{>}_{\B}
  \equiv \la \hat f_{k}(0+)\hat f_{j}(0)\ra_{\B} =\delta_{kj}\eta_{k},
\\
 \la \hat f_{j}\hat f_{k}\ra^{<}_{\B}
   \equiv \la \hat f_{j}(0)\hat f_{k}(0+)\ra_{\B}=\delta_{kj}\eta^{\ast}_{\bar k}.
\end{split}
\ee
They differ from $\la \hat f_{k}\hat f_{j}\ra_{\B}$, which will also
appear explicitly in the dissipaton algebra [cf. \Eq{Wick2}].
From \Eq{xB_in_f}, we have $\la\hat x^2_{\B}\ra_{\B} = \sum_{kj} \la \hat f_{k}\hat f_{j}\ra_{\B}$.

 As defined in \Eq{ff_t}, dissipatons $\{\hat f_k\}$
are \emph{statistically independent} quasi--particles.
Each of them is a macroscopic \emph{linear} combination of bath degrees of freedoms.
Individual $\hat f_k$ is characterized by
\emph{a single damping parameter}, $\gamma_{k}$, that can be complex,
and a \emph{joint--pair} of interacting strength parameters, $\eta_k$ and $\eta^{\ast}_{\bar k}$.

\subsection{Dissipaton algebra and DEOM formalism}
\label{thsec2B}

 Dynamical variables in DEOM are
dissipaton density operators (DDOs):\cite{Yan14054105,Yan16110306}
\be\label{DDOs}
  \rho^{(n)}_{\bf n}(t)\equiv \rho^{(n)}_{n_1\cdots n_K}(t)\equiv
   {\rm tr}_{\B}\big[
    (\hat f_{K}^{n_K}\cdots\hat f_{1}^{n_1})^{\circ}\rho_{\T}(t)\big].
\ee
The dissipatons product inside $(\cdots)^{\circ}$
is \emph{irreducible}, such that
$(\text{c-number})^{\circ}=0$.
Bosonic dissipatons satisfy the symmetric  permutation,
$(\hat f_k\hat f_j)^{\circ}=(\hat f_j\hat f_k)^{\circ}$.
%%%
Physically, each DDO of \Eq{DDOs} stands for a
given configuration of the total
$n=n_1+\cdots +n_K$ dissipatons.
%%
%%%
 Denote for the use below the associated DDO's index, ${\bf n}^{\pm}_k$,
which differs from ${\bf n}\equiv n_1\cdots n_K$ at the specified $n_k$ by $\pm 1$.
Similarly, ${\bf n}^{\pm\pm}_{kj}$ differs
from ${\bf n}$ at the specified $n_k$ and $n_j$
that are replaced by $n_k\pm 1$ and $n_j\pm 1$, respectively.

 The DEOM formalism can be easily constructed via
the algebraic dissipaton approach.\cite{Yan14054105,Yan16110306}
The construction starts  with applying the Liouville--von Neuman equation,
$\dot\rho_{\T}(t)=-i[H_{\tS}+h_{\B}+H_{\SB},\rho_{\T}(t)]$,
for the total density operator in \Eq{DDOs}.
The bath $h_{\B}$-action and the system--bath coupling $H_{\SB}$-action
are then readily evaluated with the
\emph{generalized diffusion equation} and the \emph{generalized Wick's theorem}, respectively.

  The generalized diffusion equation arises
from the single--damping parameter characteristic, as shown in \Eq{ff_t},
for both its forward and backward correlation functions.
%%%
This feature leads to \cite{Yan14054105,Yan16110306}
\be\label{diff0}
 {\rm tr}_{\B}\Big[\Big(\frac{\partial \hat f_k}{\partial t}\Big)_{\B}\rho_{\T}(t)\Big]
 = -\gamma_{k}\,{\rm tr}_{\B}\big[\hat f_k\rho_{\T}(t)\big].
\ee
Together $\big(\frac{\partial \hat f_k}{\partial t}\big)_{\B}=-i[\hat f_k, h_{\B}]$,
the generalized diffusion equation leads immediately to
\begin{align}\label{diff}
 &\quad\
  i\,{\rm tr}_{\B}\!\big\{\!
   (\hat f_{K}^{n_K}\cdots\hat f_{1}^{n_1})^{\circ} [h_{\B},\rho_{\T}(t)]\big\}
\nl&
  = i\,{\rm tr}_{\B}\!\left\{\!
    \big[(\hat f_{K}^{n_K}\cdots\hat f_{1}^{n_1})^{\circ},h_{\B}\big]\rho_{\T}(t)\right\}
\nl&
  = \Big(\sum_{k=1}^{K} n_k\gamma_k\Big)\rho^{(n)}_{\bf n}(t).
\end{align}
This is the bath $h_{\B}$--action contribution to the DDOs dynamics.
The generalized diffusion equation (\ref{diff0}) or (\ref{diff}),
with an arbitrary total composite $\rho_{\T}(t)$,
had been validated previously.\cite{Yan14054105,Yan16110306}

 The generalized Wick's theorem (GWT) deals with
the system--bath coupling $H_{\SB}$-action.\cite{Yan14054105,Yan16110306}
In this work this theorem has two ingredients,
GWT-1 and GWT-2, in relation to the linear and quadratic bath couplings,
respectively.
%%%
The GWT-1 reads\cite{Yan14054105,Yan16110306}
\begin{align}\label{Wick1}
  &\quad\
  {\rm tr}_{\B}\big[(\hat f_{K}^{n_K}\cdots\hat f_{1}^{n_1})^{\circ}
    \hat f_j\rho_{\T}(t)\big]
\nl&= \rho^{(n+1)}_{{\bf n}^{+}_{j}}(t)
     + \sum_{j} n_k \la \hat f_{k}\hat f_{j}\ra^{>}_{\B}\rho^{(n-1)}_{{\bf n}^{-}_{k}}(t).
\end{align}
%%%
The expression for
${\rm tr}_{\B}\big[(\hat f_{K}^{n_K}\cdots\hat f_{1}^{n_1})^{\circ}
\rho_{\T}(t)\hat f_j\big]$ is similar,
but goes with $\la \hat f_{j}\hat f_{k}\ra^{<}_{\B}$.
It together with
$\la \hat f_{k}\hat f_{j}\ra^{>}_{\B}$ were defined in \Eq{ff_t0}.
The GWT-1 has been well established and used in
evaluating the commutator action of linear bath coupling terms.\cite{Yan14054105,Yan16110306}

 The GWT-2 is concerned with the quadratic bath couplings,
where a pair of dissipatons $(\hat f_j\hat f_{j'})$ participate in
simultaneously without time--ordering.
This new ingredient of dissipaton algebra would reads
\bsube\label{Wick2}
\begin{align}\label{Wick20}
  &\quad\,{\rm tr}_{\B}\big[(\hat f_{K}^{n_K}\cdots\hat f_{1}^{n_1})^{\circ}
    (\hat f_j\hat f_{j'})\rho_{\T}(t)\big]
%%%
\nl&= \sum_{k} n_k \la\hat f_k\hat f_j\ra^{>}_{\B}
   {\rm tr}_{\B}\big[(\hat f_{K}^{n_K}\cdots \hat f^{n_k-1}_k\cdots
     \hat f_{1}^{n_1})^{\circ}\hat f_{j'}\rho_{\T}(t)\big]
\nl&\quad
 +{\rm tr}_{\B}\big[(\hat f_{K}^{n_K}\cdots\hat f_{1}^{n_1}\hat f_j)^{\circ}
    \hat f_{j'}\rho_{\T}(t)\big],
%%%
\end{align}
with the last term being evaluated as
\begin{align}\label{Wick21}
  &\quad\ {\rm tr}_{\B}\big[(\hat f_{K}^{n_K}\cdots\hat f_{1}^{n_1}\hat f_j)^{\circ}
    \hat f_{j'}\rho_{\T}(t)\big]
\nl&=
   \rho^{(n+2)}_{{\bf n}^{++}_{jj'}}(t)
  + \la\hat f_j \hat f_{j'}\ra_{\B}\rho^{(n)}_{{\bf n}}(t)
\nl&\quad\,
  + \sum_{k} n_k \la\hat f_k\hat f_{j'}\ra^{>}_{\B}\rho^{(n)}_{{\bf n}^{-+}_{kj}}(t).
\end{align}
\esube
%%%
Together with the first term in \Eq{Wick20} being
evaluated via GWT-1, we obtain
%%%
\begin{align}\label{Wick2_all}
%%%
 &\quad\,{\rm tr}_{\B}\big[(\hat f_{K}^{n_K}\cdots\hat f_{1}^{n_1})^{\circ}
    (\hat f_j\hat f_{j'})\rho_{\T}(t)\big]
\nl&= \rho^{(n+2)}_{{\bf n}^{++}_{jj'}}(t) + \la\hat f_j \hat f_{j'}\ra_{\B}\rho^{(n)}_{{\bf n}}(t)
\nl&\quad
  +\sum_{k} n_k \big[
      \la\hat f_k\hat f_j\ra^{>}_{\B}\rho^{(n)}_{{\bf n}^{-+}_{kj'}}(t)
     +\la\hat f_k\hat f_{j'}\ra^{>}_{\B}\rho^{(n)}_{{\bf n}^{-+}_{kj}}(t)
   \big]
\nl&\quad
  + \sum_{k,k'} n_k(n_{k'}-\delta_{kk'})\la\hat f_k\hat f_j\ra^{>}_{\B}\la\hat f_{k'}\hat f_{j'}\ra^{>}_{\B}
    \rho^{(n-2)}_{{\bf n}^{--}_{k k'}}(t).
\end{align}
The expression for
${\rm tr}_{\B}\big[(\hat f_{K}^{n_K}\cdots\hat f_{1}^{n_1})^{\circ}
\rho_{\T}(t)(\hat f_j\hat f_{j'})\big]$ is similar,
but with each individual $\la \hat f_{k}\hat f_{j}\ra^{<}_{\B}$
being replaced by $\la \hat f_{j}\hat f_{k}\ra^{<}_{\B}$.
%%%
The GWT-2 is to be used in the evaluation of
the quadratic bath coupling contribution.

 The DEOM formalism in the presence
of both linear and quadratic bath couplings
can now be readily constructed via the above dissipaton algebra.
The final results read
\begin{align}\label{DEOM}
  \dot\rho^{(n)}_{\bf n} \!
& = - \Big(i{\cal L}_{\rm eff} + \sum_{k}n_k\gamma_k\Big)\rho^{(n)}_{\bf n}
  -i2\alpha_2 \sum_{kj} n_k {\cal C}_k \rho^{(n)}_{{\bf n}^{-+}_{kj}}
\nl&\quad
  -i\alpha_2\sum_{kj} \Big[{\cal A}\rho^{(n+2)}_{{\bf n}^{++}_{kj}}
    +n_k(n_{j}-\delta_{jk}){\cal B}_{kj}\rho^{(n-2)}_{{\bf n}^{--}_{kj}}
 \Big]
\nl&\quad
  -i\alpha_1\sum_{k}\Big({\cal A}\rho^{(n+1)}_{{\bf n}^{+}_{k}}
      +n_k{\cal C}_k\rho^{(n-1)}_{{\bf n}^{-}_{k}}\Big).
\end{align}
Here,
\bsube\label{calABC}
\begin{align}\label{calHA}
  {\cal L}_{\rm eff}\hat O &\equiv [H_{\rm eff},\hat O],
\ \ \,
  {\cal A}\hat O \equiv \big[\hat Q_{\tS}, \hat O],
%%%
\\  \label{calB}
  {\cal B}_{kj} \hat O &\equiv  \eta_k\eta_j \hat Q_{\tS} \hat O
     - \eta^{\ast}_{\bar k}\eta^{\ast}_{\bar j}\hat O \hat Q_{\tS},
%%%
\\ \label{calB}
  {\cal C}_k\hat O &\equiv \eta_k \hat Q_{\tS}\hat O
        -\eta^{\ast}_{\bar k}\hat O\hat Q_{\tS},
\end{align}
\esube
with
\be\label{Heff}
  H_{\rm eff} \equiv H_{\tS}+\alpha_2\la\hat x^2_{\B}\ra_{\B}\hat Q_{\tS}.
\ee
%%%

Evidently, the dissipatons defined in \Eqs{FDT}--(\ref{ff_t})
are strictly based on the linear bath coupling part that satisfies
Gaussian--Wick's statistics.\cite{Wei08,Kle09,Yan05187}
The quadratic non--Gaussian bath influences
are treated via the GWT-2, \Eq{Wick2_all}.
This is the virtue of the DEOM theory
that includes the powerful dissipaton algebra.\cite{Yan14054105,Yan16110306}
In general, \Eqs{diff0}--(\ref{Wick2_all}) are all
non--Gaussian operators in the system subspace.
%%%

\section{Validation on dissipaton algebra with extended Zusman equation}
\label{thsec3}

 This section is devoted to validate the GWT-2, \Eq{Wick2},
the new ingredient of the dissipaton algebra presented in \Sec{thsec2}.
It together with the well--established \Eqs{diff0} and (\ref{Wick1})
lead immediately and unambiguously to the extended DEOM (\ref{DEOM}).
%%%
Therefore, from the algebraic construction point of view,
the required validation can be made
with the dissipaton basis set of size $K=1$.
%%%
This amounts to the formal setting \Eq{FDT} with
\be\label{xxt_single}
 \la \hat x_{\B}(t)\hat x_{\B}(0)\ra_{\B} \simeq \eta e^{-\gamma t}.
\ee
The dissipaton index ${\bf n}$ can be omitted; i.e.,
$\rho^{(n)}_{\bf n}(t)= \rho^{(n)}(t)\equiv \hat\rho_{n}(t)$,
for the basis set size $K=1$ case, in which
%%%
the DEOM (\ref{DEOM}) reads
\begin{align}\label{DEOM_single}
 \dot{\hat\rho}_{n}(t)
& = - ( i{\cal L}_{\rm eff} + n\gamma +2\alpha_2n{\cal C}) \hat\rho_{n}(t)
\nl&\quad
  -i\alpha_2\left[{\cal A}\hat \rho_{n+2}(t)+n(n-1){\cal B}\hat\rho_{n-2}(t)
           \right]
\nl&\quad
  -i\alpha_1\left[{\cal A}\hat \rho_{n+1}(t) + n{\cal C}\hat \rho_{n-1}(t) \right].
\end{align}
%%%
The involving superoperators ${\cal A}$, ${\cal B}$ and ${\cal C}$,
are the same as those in \Eq{calABC}, but without dissipaton indexes.

 In the following, on the basis of \Eq{xxt_single} which will be
called the Zusman setup, we construct the extended Zusman equation via a totally different approach.
By showing that the extended Zusman equation is identical to \Eq{DEOM_single},
we validate the dissipaton algebra and thus the
extended DEOM theory, \Eq{DEOM}.
The proof here is \emph{rigourous}, due to the nature of the algebraic construction,
despite of the fact that the Zusman setup, \Eq{xxt_single}, itself
could even be a bad approximation.

 It is well known that the Zusman setup, \Eq{xxt_single},
is equivalent to the combination of
the high--temperature (HT) and the Smoluchowski limits.
The HT limit is characterized with
\be\label{higtT}
 \frac{1}{1-e^{-\beta\omega}} \simeq \frac{1}{\beta\omega}+\frac{1}{2}.
\ee
In this case, the solvation mode is a classical Brownian motion in the secondary bath environment.
The latter exerts a stochastic force $\wti F(t)$ and
friction constant $\zeta$ on the solvation mode.
The corresponding Langevin equation reads
\be\label{Langevin}
 \dot{p}_{\B}(t)=-\w_{\B}x_{\B}(t)
  -\zeta p_{\B}(t) +\wti F(t),
\ee
with
\be\label{corr_HT}
   \la \wti F(t)\wti F(0)\ra^{\text{\tiny HT}} = \frac{2\zeta}{\beta\w_{\B}}\delta(t)
  + i \frac{\zeta}{\w_{\B}}\,{\dot\delta}(t).
\ee
This is the high--temperature fluctuation--dissipation theorem.
The resultant Caldeira--Leggett's equation reads \cite{Cal83587,Gar854491}
\begin{align}\label{rhoCL}
 \frac{\partial}{\partial t}{\hat\rho}^{\text{\tiny HT}}_{\text{\tiny W}}
 & =-(i{\cal L}_{\tS}+\hat L_{\text{\tiny FP}}){\hat\rho}^{\text{\tiny HT}}_{\text{\tiny W}}
    +
    \frac{\partial}{\partial p_{\B}}\Big(\frac{\alpha_1}{2}
    +\alpha_2x_{\B}\Big)
   \{\hat Q_{\tS},
    {\hat\rho}^{\text{\tiny HT}}_{\text{\tiny W}}
   \}
 \nl & \quad
   -i\Big(\alpha_1x_{\B}
     +\alpha_2x_{\B}^2-\frac{\alpha_2}{4} \frac{\partial^2}{\partial p_{\B}^2}\Big)
     [\hat Q_{\tS},
     {\hat\rho}^{\text{\tiny HT}}_{\text{\tiny W}}].
\end{align}
Here, ${\hat\rho}^{\text{\tiny HT}}_{\text{\tiny W}}
\equiv {\hat\rho}^{\text{\tiny HT}}_{\text{\tiny W}}(x_{\B},p_{\B};t)$
denotes the reduced system--plus--solvation density operator in the HT limit,
with the solvation mode in the Wigner representation.
While ${\cal L}_{\tS}\,\cdot\, \equiv [H_{\tS},\,\cdot\,]$ is the system Liouvillian,
$\hat L_{\text{\tiny FP}}$ in \Eq{rhoCL} is the Fokker--Planck operator,\cite{Ris89}
\be\label{calFP}
  \hat L_{\text{\tiny FP}}=
  \w_{\B}\Big(\frac{\partial}{\partial x_{\B}}p_{\B}
 -\frac{\partial}{\partial p_{\B}}x_{\B}\Big)
 - \frac{\zeta}{\beta\w_{\B}}
   \frac{\partial^2}{\partial p_{\B}^2}
 -\zeta\frac{\partial}{\partial p_{\B}}p_{\B} .
\ee

 To complete the Zusman setup, \Eq{xxt_single},
consider further the Smoluchowski (or strongly--overdamped) limit;
i.e., $\zeta\gg \w_{\B}$,
whereas $\w_{\B}^2/\zeta=\gamma$ remains finite to
be the exponent in \Eq{xxt_single}.
Moreover, it is easy to show that
$\la x^2_{\B}\ra_{\B} = \la p^2_{\B}\ra_{\B}\rightarrow(\beta\w_{\B})^{-1}$,
in the high--temperature limit.
The identities here will be used in
eliminating the appearance of $\beta\w_{\B}$
in the formulations below.
The pre-exponential coefficient in \Eq{xxt_single}
reads then
\be\label{eta_Zusman}
   \eta \equiv \eta_r+i\eta_i = \la x^2_{\B}\ra_{\B}(1-i\beta\gamma/2) .
\ee
In the strongly--overdamped limit,
the momentum $p_{\B}$ would no longer be a correlated dynamical variable.
The equation of motion for
$\hat\rho(x_{\B};t)=\int{\rm d}p_{\B}\hat\rho_\text{\tiny W}(x_{\B},p_{\B};t)$,
which is closed now,
can be obtained via the standard Fokker--Planck--Smoluchowski algorithm.\cite{Ris89}

 Another equivalent but much simpler approach is
the so--called diffusion mapping method;\cite{Zha14319}
i.e., mapping each individual $p_{\B}$--space variable to
its limiting diffusive $x_{\B}$--space correspondence.
This method makes a simple use of the Langevin equation (\ref{Langevin}), which
in the strongly--overdamped limit reduces to
\be\label{Smolimit0}
  0 = -\dot p_{\B} = \omega_{\B} x_{\B} + \zeta p_{\B} - \wti F(t).
\ee
%%%
 Consider further the following two thermodynamic relations,
\be\label{partial_xp}
\begin{split}
 \frac{\partial}{\partial p_{\B}}\hat\rho^{\text{\tiny HT}}_{\text{\tiny W}}
 &= -\frac{1}{\la p_{\B}^2\ra}_{\B} p_{\B} \hat\rho^{\text{\tiny HT}}_{\text{\tiny W}},
\\
 \frac{\partial}{\partial x_{\B}}\hat\rho^{\text{\tiny HT}}_{\text{\tiny W}}
 &= -\frac{1}{\la x_{\B}^2\ra}_{\B} x_{\B} \hat\rho^{\text{\tiny HT}}_{\text{\tiny W}}.
%%%
\end{split}
\ee
Together with $p_{\B}/x_{\B} \simeq -\omega_{\B}/\zeta$,
as implied in \Eq{Smolimit0}, and ${\la x_{\B}^2\ra}_{\B}={\la p_{\B}^2\ra}_{\B}$
in study here,
we would have then
%%%
\be\label{map0pm}
\frac{\partial}{\partial p_{\B}} \hat\rho^{\text{\tiny HT}}_{\text{\tiny W}}
 \simeq \frac{\omega_{\B}}{\zeta}
  \frac{\partial}{\partial x_{\B}}\hat\rho^{\text{\tiny HT}}_{\text{\tiny W}} \, .
\ee
The above results of the Zusman setup lead to the
following rules of diffusion mapping,\cite{Zha14319}
%%%%
\be\label{highFriction_map}
  \frac{\partial}{\partial p_{\B}}
   \Longrightarrow
   \frac{\omega_{\B}}{\zeta}\frac{\partial}{\partial x_{\B}}
\ \ \text{and} \ \
  p_{\B} \Longrightarrow  -{\la x_{\B}^2\ra}_{\B}\frac{\omega_{\B}}{\zeta}
       \frac{\partial}{\partial x_{\B}}.
\ee
%%%
%%%%
The Smoluchowski limit to the Calderia--Leggett's master equation is now readily
obtained by replacing all those $p_{\B}$-dependent operators
in \Eqs{rhoCL} and (\ref{calFP}).
In particular, the Fokker--Planck operator becomes the
Smoluchowski or diffusion operator,
\be\label{LD}
 {\hat L}_{\text{\tiny FP}}  \Longrightarrow
 {\hat L}_{\text{\tiny D}}
\equiv
 -\gamma \Big(\eta_r\frac{\partial^2}{\partial x^2_{\B}}
              + x_{\B}\frac{\partial}{\partial x_{\B}}\Big).
\ee
While $\eta_r={\la x_{\B}^2\ra}_{\B}$ was defined in \Eq{eta_Zusman},
%%%%
$\gamma = \omega^2_{\B}/\zeta$ assumes the diffusion constant here.
%%%
The Smoluchowski limit of \Eq{rhoCL} has
an extended Zusman equation form,
%%%
\begin{align}\label{exZE1}
 &\quad\, \frac{\partial}{\partial t}\hat\rho(x_{\B};t)
+ (i{\cal L}_{\tS}+{\hat L}_{\D})\hat\rho(x_{\B};t)
%%%%%%
\nl&=
   -i\Big[\alpha_1 x_{\B}+\alpha_2\Big(x_{\B}^2
   -\eta^{2}_i \frac{\partial^2}{\partial x_{\B}^2}\Big)\Big]
    \big[\hat Q_{\tS},\hat\rho(x_{\B};t)\big]
\nl&\quad
  - 2\eta_i \frac{\partial}{\partial x_{\B}}
    \left[\big(\frac{\alpha_1}{2}+\alpha_2 x_{\B}\big)
      \big\{\hat Q_{\tS},\hat\rho(x_{\B};t)\big\}
    \right].
\end{align}
%%%%
This recovers the conventional Zusman equation,\cite{Zus80295,Zus8329,Gar854491,Yan89281}
in the absence of the quadratic bath coupling ($\alpha_2=0$).

 We have also derived \Eq{exZE1} via the standard Fokker--Planck--Smoluchowski
approach;\cite{Ris89} however, the derivations are too mathematical and tedious,
see Appendix for details.
The universal diffusion mapping approach with \Eq{highFriction_map}
is much simpler and physically more appealing.

 It is easy to verify that the DEOM formalism, \Eq{DEOM_single},
is just the dynamical resolution to the extended Zusman equation (\ref{exZE1}).
More precisely,
\be\label{rhon_rho_x}
  \hat\rho_n(t)=  \Big(\frac{\eta_r}{2}\Big)^{\!\frac{n}{2}}
  \!\!\int_{-\infty}^{\infty}\!\!{\rm d}x_{\B}
  H_n\Big(\frac{x_{\B}}{\sqrt{2\eta_r}}\Big)\hat\rho(x_{\B};t),
\ee
or
\be\label{tirhoexpan}
  \hat\rho(x_{\B};t) = \sum_{n=0}^{\infty} \phi_n(x_{\B})\hat\rho_n(t) ,
\ee
where $H_n(x)$ is the $n^{\rm th}$-order Hermite polynomial, and
%%%
\be\label{phin}
  \phi_n(x_{\B})=
  \frac{(2\eta_r)^{-\frac{n}{2}}
       }{n!\sqrt{2\pi\eta_r}}
  \exp\Big(\!-\!\dfrac{x^2_{\B}}{2\eta_r}\Big)
    H_n\Big(\dfrac{x_{\B}}{\sqrt{2\eta_r}}\Big).
\ee

 We have thus validated the dissipaton algebra, \Eqs{diff0}--(\ref{Wick2_all}).
%%%
This is the purpose of the above comparisons
between the dissipaton approach in \Sec{thsec2} and the
present system--and--solvation composite
description.
%%%%
The dissipaton algebra, including the new ingredient, \Eq{Wick2}, the generalized
Wick's theorem with dissipatons--pairs added,
is also by \emph{de facto} established.
%%%
Therefore, the extended DEOM (\ref{DEOM}) for general
cases is also validated, due to its algebraic construction nature.

\section{Interplay between dissipatons and environment parameters}
\label{thsec4}

  Turn to the issue on the bath coupling parameters, $\alpha_1$ and $\alpha_2$.
It is crucial to have a physical support on the nonlinear
coupling bath descriptors. This issue is directly related to
the extended DEOM theory, which
should describe the dynamical interplay between dissipatons
and nonlinear bath couplings. Erroneous descriptors
of $\alpha_1$ and $\alpha_2$
would result in unphysical DEOM dynamics.

  In the following, we adopt a polarization model\cite{Xu17JCP1}
to determine both $\alpha_1$ and $\alpha_2$.
For clarity, we consider a chromophore system, with
its ground $|g\ra$ and an excited $|e\ra$ states
being engaged in optical excitations.
%%%
The total system--and--bath composite
Hamiltonian in the presence of external classical laser field $E(t)$
assumes
\be\label{HT_spectrum}
 H_{\T}(t)=h_g|g\ra\la g|+(h_e+\w_{eg})|e\ra\la e| - \hat \mu_{\tS} E(t),
\ee
with $\hat \mu_{\tS} = \mu(|g\ra\la e|+|g\ra\la e|)$ and
\be\label{delta_hB}
  h_e-h_g = \alpha_0 + \alpha_1 \hat x_{\B} + \alpha_2\hat x^2_{\B}.
\ee
Here, $h_g$ and $h_e$ denote the bath Hamiltonians associating
with the ground and excited system states, respectively.
Equation (\ref{HT_spectrum}) assumes the form of \Eq{HT}; i.e.,
\be\label{HT1}
 H_{\T}(t)  =  H_{\tS}(t) + h_{\B} + \hat Q_{\tS}(\alpha_1\hat x_{\B}+\alpha_2\hat x^2_{\B}),
\ee
with the system Hamiltonian and dissipative mode,
\be\label{HQS}
   H_{\tS}(t)=(\w_{eg}+\alpha_0)|e\ra\la e|-\hat\mu_{\tS}E(t),
\ \ \,
    \hat Q_{\tS}=|e\ra\la e|.
\ee
In \Eq{HT1} the bath Hamiltonian goes with $h_{\B}=h_g$.

The polarization model assumes\cite{Xu17JCP1}
\be\label{hB_solvA}
\begin{split}
  h_{g}  &= \frac{1}{2}\w_{\B}(\hat p^2_{\B}+\hat x^2_{\B})
    + h^{\rm int}_{\B}(\hat x_{\B};\ti{\bf q}),
\\
  h_{e}  &= \frac{1}{2}\w'_{\B}(\hat p^{\prime 2}_{\B}+\hat x^{\prime 2}_{\B})
    + h^{\rm int}_{\B}(\hat x'_{\B};\ti{\bf q}),
\end{split}
\ee
where $\hat p'_{\B}=(\w_{\B}/\w'_{\B})^{\frac{1}{2}}\hat p_{\B}$,
\be\label{xB_e}
 \hat x'_{\B} = (\w'_{\B}/\w_{\B})^{1/2}(\hat x_{\B}-d_{\B}) ,
\ee
and
\be\label{hB_int_def}
  h^{\rm int}_{\B}(\hat x_{\B};\ti{\bf q})
   =\frac{1}{2}\sum_{k}\ti\w_k \big[
     \ti p^2_k+(\ti q_k-\ti c_k\hat x_{\B})^2
    \big].
\ee
%%%
The physical picture of this model is as follows.
The system is initial in the ground $|g\ra$ state,
with $\hat x_{\B}$ describing
its first solvation shell of frequency $\w_{\B}$.
Upon excitation, the system in the excited $|e\ra$ state
experiences different solvation environment.
The reorganized first--shell solvation is described with $\hat x'_{\B}$.
It has different frequency, $\w'_{\B}$, and is also linearly shifted
by $d_{\B}$, with respectively to the ground--state solvation shell.
%%%
The secondary environment ($\ti{\bf q}$) remains unchanged, as described
by the same $h^{\rm int}_{\B}(X_{\B};\ti{\bf q})$,
for  its interacting with  either $X_{\B}=\hat x_{\B}$
or $\hat x'_{\B}$  solvation mode.
Apparently, the quadractic bath coupling vanishes when $\w'_{\B}=\w_{\B}$.

 The coupling strength of secondary bath with the solvation mode is given by
%%%%
\be\label{wti_eta}
 \wti\eta \equiv \sum_{k} \ti\w_k\ti c^2_k.
\ee
The renormalized frequencies for $\hat x_{\B}$ and $\hat x'_{\B}$ would be
\be\label{wti_wB}
  \wti\w_{\B}=\w_{\B}+\wti\eta
\ \ {\rm and} \ \
  \wti\w'_{\B} = \w'_{\B} + \wti\eta,
\ee
respectively. The reorganization energy in the absence of quadratic bath coupling
is given by
\be\label{lambda}
  \lambda \equiv \frac{1}{2}\wti\w_{\B}d^2_{\B}.
\ee
 Substituting \Eqs{hB_solvA}--(\ref{lambda}) for \Eq{delta_hB},
followed by some algebra, we obtain\cite{Xu17JCP1}
\be\label{alpha_all}
\begin{split}
 \alpha_0 &= \theta\ti\theta\lambda ,
\ \ \
 \alpha_1 = \sqrt{2\theta\lambda\w_{\B}}\Big(1+\frac{\sqrt{\theta}\ti\theta-1}{\varphi}\Big),
%%%%
\\
 \alpha_2 &=\frac{\wti\w_{\B}}{2} \left[2(\sqrt{\theta}-1)\varphi
   +(\theta\ti\theta -2\sqrt{\theta}+1)
   \right] ,
\end{split}
\ee
where
\be\label{theta_phi}
  \theta \equiv \frac{\w'_{\B}}{\w_{\B}},
\ \ \,
  \wti\theta \equiv  \frac{\wti\w'_{\B}}{\wti\w_{\B}},
\ \ \,
  \varphi \equiv \sqrt{ \frac{\w_{\B}}{\wti\w_{\B}} }.
\ee
Apparently, $\alpha_2= 0$ if $\w'_{\B}= \w_{\B}$.

 Moreover, the bath coupling $\alpha$--parameters of \Eq{alpha_all}
should go along with the underlying interplay with the dissipatons
defined in \Eqs{FDT}--(\ref{ff_t}).
%%%
Given the temperature, dissipatons are
determined by $\chi^{(i)}_{\B}(\w)$.
%%%
On the other hand, the $\alpha$--parameters in \Eq{alpha_all}
are functions of $\wti\eta$, $\w_{\B}$, $\w'_{\B}$ and $\lambda$.
While the latter two are free variables, $\wti\eta$ and $\w_{\B}$
are dictated by the same $\chi^{(i)}_{\B}(\w)$
that determines the dissipatons.
%%%%
%
 For the secondary bath coupling strength parameter [\Eq{wti_eta}],
we have \cite{Yan05187,Xu17JCP1}
\be\label{wti_eta_wtiJ}
 \wti\eta
 = \frac{1}{\pi}\int^{\infty}_{-\infty}\!\! {\rm d}\w \frac{\wti J(\w)}{\w}.
\ee
Here, $\wti J(\w)$ denotes the interacting secondary bath spectral density that
can be expressed in terms of \cite{Yan05187}
\be\label{tiJw_in_chiBw}
  \wti J (\w) = \frac{\chi^{(i)}_{\B}(\w)}{|\chi_{\B}(\w)|^2}.
\ee
Note that $\chi_{\B}(\w)\equiv \chi^{(r)}_{\B}(\w)+i\chi^{(i)}_{\B}(\w)$
is completely determined by either the real or imaginary part
via the Kramers--Kronig relation.\cite{Wei08,Kle09,Yan05187}
 For the solvation mode frequency, we have\cite{Yan05187,Xu17JCP1}
\be\label{wB_via_chi}
 \w_{\B} = \frac{1}{\chi_{\B}(0)}
  = \frac{1}{\pi}\int^{\infty}_{-\infty}\!\! {\rm d}\w\, \w \chi^{(i)}_{\B}(\w).
\ee
Here, $\chi_{\B}(0)=\chi^{(r)}_{\B}(0)$, as $\chi^{(i)}_{\B}(\w=0)=0$.
The Kramers--Kronig relation reads here\cite{Wei08,Kle09,Yan05187}
\be\label{KK0}
 \chi_{\B}(0) = \frac{1}{\pi}\int^{\infty}_{-\infty}\!\!
    {\rm d}\w\, \frac{\chi^{(i)}_{\B}(\w)}{\w}.
\ee
The above identities describe the determination
of $\w_{\B}$ via any given $\chi^{(ii)}_{\B}(\w)$.
Actually, \Eq{wB_via_chi} follows in line with
the definition of $\chi_{\B}(\w)$ in \Eq{chiw_def}.

\begin{figure}
\includegraphics[width=0.75\columnwidth]{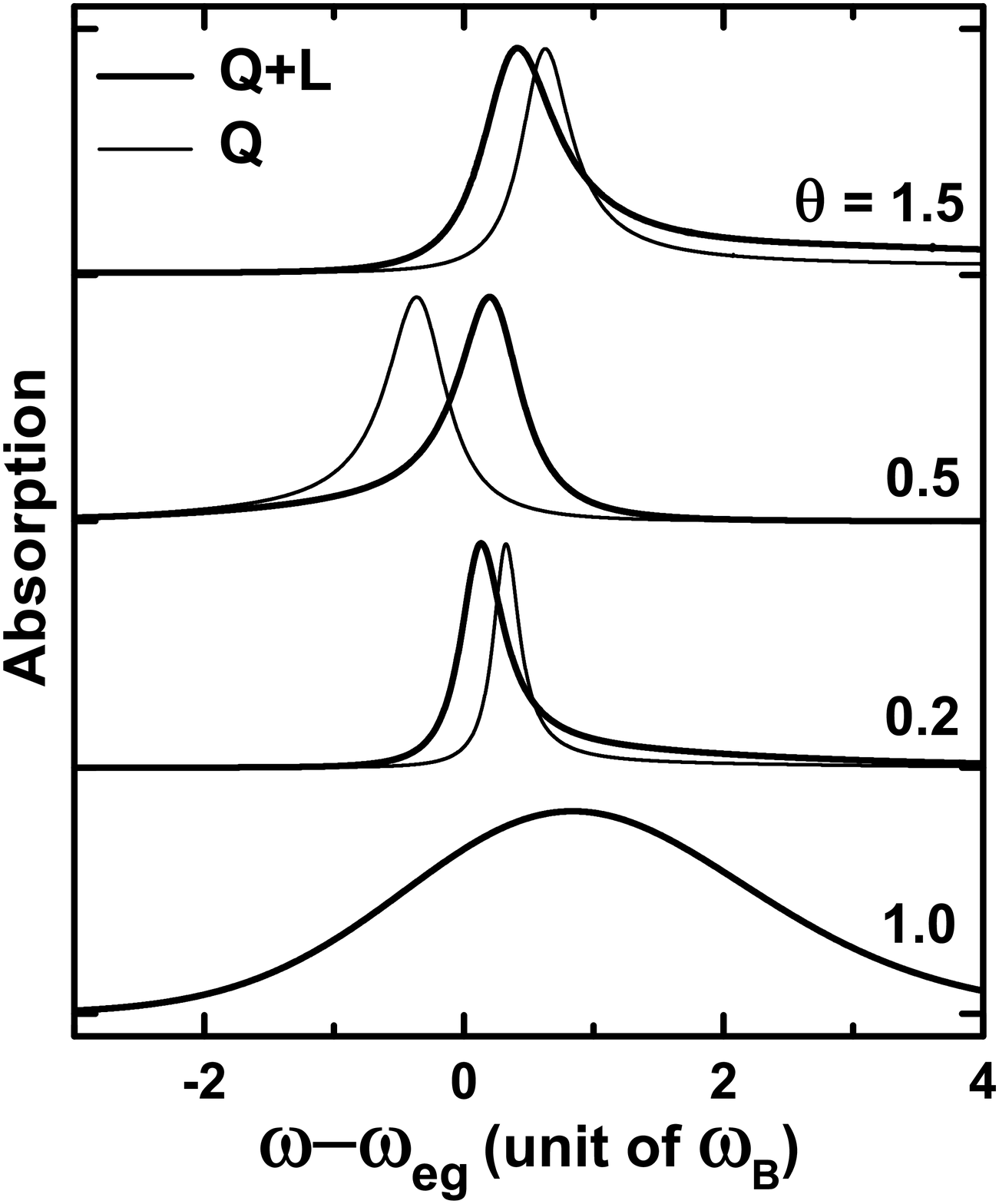}
\caption{Evaluated absorption lineshapes. See text for the details.
}
\label{fig1}
\end{figure}

  For numerical illustrations, we adopt the Drude model
for the secondary bath. The resultant $\chi_{\B}(\w)$,
which supports \Eqs{wti_eta_wtiJ}--(\ref{KK0}),
reads
\be\label{chiBw_model}
 \chi_{\B}(\w) = \frac{\w_{\B}}{
       \w^2_{\B}-\w^2-\dfrac{\wti\eta\w_{\B}\w}{\w+i\wti\gamma}
      }.
\ee
We set the parameters (in unit of $\w_{\B}$)
$\wti\gamma=10$ and $\wti\eta=20$, at $k_{B}T=1$;
and also $\lambda=1$ for nonzero linear bath coupling strength.
%%%
Figure \ref{fig1} depicts the evaluated linear absorption
spectrums, at four representing values of $\theta=\w'_{\B}/\w_{\B}$.
%%%
When $\theta=1$ there is only the linear bath coupling.
The other three (solid) curves with $\theta\neq 1$
are of both the linear and the quadratic bath couplings,
which are presented in parallel with
their linear--free (thin) counterparts.
%%%%
In contrast with the pure linear bath coupling ($\theta=1$) case,
the spectrum lineshape is generally asymmetric.
%%%%
The observed skews in individual lineshape profiles, which show non-monotonic dependence
on $\theta$, are all in qualitative agreements
with the secondary--bath--free but analytical results.\cite{Yan865908}

\section{Summary}
\label{thconc}

 Evidently, the dissipaton algebra leads readily to
the DEOM formalism, as seen in \Sec{thsec2}.
%%%
The key contribution of this work is the establishment
the generalized Wick's theorem with a pair of dissipatons added;
i.e., the GWT-2, \Eq{Wick2}.
The other ingredients of the dissipaton algebra presented in \Sec{thsec2}
had all been well established in our previous work.\cite{Yan14054105,Yan16110306}

 The new ingredient, \Eq{Wick2}, which is now verified unambiguously
in \Sec{thsec3},
can be used consecutively to treat further
\emph{higher--order} nonlinear bath couplings.
%%%
For example, the GWT-3, illustrated with the dissipaton--basis--set size of one,
would go with [cf.\ \Eq{Wick20}]
\begin{align*}
  {\rm tr}_{\B}\big[(\hat f^n)^{\circ}(\hat f \hat f \hat f)\rho_{\T}\big]
&=n\la\hat f\hat f\ra^{>}_{\B} {\rm tr}_{\B}\big[(\hat f^n)^{\circ}(\hat f \hat f)\rho_{\T}\big]
\nl&\quad
 +{\rm tr}_{\B}\big[(\hat f^n\hat f)^{\circ}(\hat f \hat f)\rho_{\T}\big].
\end{align*}
While the first quantity is evaluated by using \Eq{Wick2},
the second quantity would be
\begin{align*}
  {\rm tr}_{\B}\big[(\hat f^n\hat f)^{\circ}(\hat f \hat f)\rho_{\T}\big]
&= \la \hat f^2\ra_{\B}{\rm tr}_{\B}\big[(\hat f^n)^{\circ}\hat f\rho_{\T}\big]
\nl&\quad
  +n\la\hat f\hat f\ra^{>}_{\B}{\rm tr}_{\B}\big[(\hat f^n\hat f)^{\circ}\hat f\rho_{\T}\big]
\nl&\quad
+{\rm tr}_{\B}\big[(\hat f^n\hat f\hat f)^{\circ}\hat f\rho_{\T}\big].
\end{align*}
The first two quantities are evaluated via the GWT-1
and the GWT-2 of \Eq{Wick21}, respectively.
The last quantity above goes with
\[
 {\rm tr}_{\B}\big[(\hat f^n\hat f\hat f)^{\circ}\hat f\rho_{\T}\big]
 = \rho^{(n+3)}+(2\la\hat f^2\ra_{\B}+n\la\hat f\hat f\ra^{>}_{\B})\rho^{(n+1)}.
\]
The GWT-3 is then completed. The GWT-n follows the same recursive procedure.
%%%%
Thus, the present work represents a major advancement
in the DEOM theory, with the specified class of non--Gaussian coupling environments
that could be physically characterized, as illustrated in \Sec{thsec4}.

\begin{acknowledgements}

  The support from
the Ministry of Science and Technology of China (Nos.\ 2017YFA0204904 \& 2016YFA0400904),
the Natural Science Foundation of China
(Nos.\ 21633006 \& 21373191),
and the Fundamental Research Funds for Central Universities (No.\ 2030020028)
is gratefully acknowledged.

\end{acknowledgements}

\begin{appendix}*

\section{The Smoluchowski limit: Conventional approach}
\label{thappA}

 This appendix utilizes the standard textbook approach of Ref.\ \onlinecite{Ris89}
to derive the extended Zusman equation (\ref{exZE1}).
This is the application of the Smoluchowski limit to
the Caldeira--Leggett's equation (\ref{rhoCL}),
and derive the closed equation for
\be\label{rhoZusman}
 \hat\rho(x_{\B};t)=\int_{-\infty}^\infty\!\! {\rm d} p_{\B}\,
    {\hat\rho}^{\text{\tiny HT}}_{\text{\tiny W}}(x_{\B},p_{\B};t).
\ee
We start with the FP operator, \Eq{calFP}, which
has the coherent and incoherent contributions:
\be\label{hatL_FP_app}
  {\hat L}_{\FP}={\hat L}^{\text{coh}}_{\FP}+{\hat L}^{\text{incoh}}_{\FP},
\ee
where
\be\label{hatL_coh_inc}
\begin{split}
  {\hat L}^{\text{coh}}_{\FP}&= \w_{\B}\Big(\frac{\partial}{\partial x_{\B}}p_{\B}
 -\frac{\partial}{\partial p_{\B}}x_{\B}\Big),
\\
  {\hat L}^{\text{incoh}}_{\FP}&=-\zeta\frac{\partial}{\partial p_{\B}}
    \Big(\frac{1}{\beta\w_{\B}}\frac{\partial}{\partial p_{\B}}+p_{\B}
    \Big).
\end{split}
\ee
It is easy to obtain\cite{Ris89}
\be\label{wtiL_FP}
 {\wti L}^{\text{incoh}}_{\FP} \equiv
   \psi^{-1}_0(p_{\B}){\hat L}^{\text{incoh}}_{\FP}\psi_0(p_{\B})
 =\zeta \hat a^{\dg}\hat a,
\ee
where
\be\label{psi_0}
  \psi_0(p_{\B})=\Big(\frac{\beta\w_{\B}}{2\pi}\Big)^{\frac{1}{4}}
      \exp\Big(-\frac{\beta\w_{\B}}{4}p_{\B}^2\Big),
\ee
and
\be\label{wtiL_FP_aa_dg}
\begin{split}
  \hat a&= \frac{\sqrt{\beta\w_{\B}}}{2} p_{\B}
          +\frac{1}{\sqrt{\beta\w_{\B}}}\frac{\partial}{\partial p_{\B}},
\\
 \hat a^{\dg}&= \frac{\sqrt{\beta\w_{\B}}}{2} p_{\B}
          -\frac{1}{\sqrt{\beta\w_{\B}}}\frac{\partial}{\partial p_{\B}}.
\end{split}
\ee
These are the bosonic annihilation and creation operators, satisfying $[\hat a,\hat a^{\dg}]=1$.
%%%%
The normalized eigen solutions to \Eq{wtiL_FP} are therefore
\be
  {\wti L}^{\text{incoh}}_{\FP}\psi_{n}(p_{\B})=n\psi_{n}(p_{\B}),
\ \ \,n=0,1,\cdots,
\ee
with the ground state $\psi_{0}(p_{\B})$ of \Eq{psi_0} and
\be\label{psi_n}
  \hat a^{\dg}\psi_{n}(p_{\B})= \sqrt{n+1}\,\psi_{n+1}(p_{\B}).
\ee
Apparently, $\{\psi_n(p_{\B})\}$ are all real.

 On the other hand, from \Eq{wtiL_FP_aa_dg}, we evaluate
\be
\begin{split}
 \wti{\frac{\partial}{\partial p_{\B}}}
&\equiv
   \psi^{-1}_0(p_{\B})\frac{\partial}{\partial p_{\B}} \psi_0(p_{\B})
 =-\sqrt{\beta\w_{\B}}\hat a^{\dg},
\\
 {\wti p}_{\B} &\equiv p_{\B} = \frac{1}{\sqrt{\beta\w_{\B}}}(\hat a+\hat a^{\dg}).
\end{split}
\ee
The transformed
${\wti L}^{\text{coh}}_{\FP}\equiv \psi^{-1}_0(p_{\B}) {\hat L}^{\text{coh}}_{\FP}  \psi_0(p_{\B})$
in \Eq{hatL_coh_inc} reads then
\begin{align} {\wti L}^{\text{coh}}_{\FP}
 = \w_{\B}\Big[
     \frac{1}{\sqrt{\beta\w_{\B}}}\frac{\partial}{\partial x_{\B}}(\hat a+\hat a^{\dg})
   + \sqrt{\beta\w_{\B}} x_{\B}\hat a^{\dg}
 \Big].
\end{align}
We have also
\be\label{wti_partial_p2}
  \wti{\frac{\partial^2}{\partial p_{\B}^2}}=\Big(\wti{\frac{\partial}{\partial p_{\B}}}\Big)^2
  =\beta\w_{\B}\hat a^{\dg\,2}.
\ee

 Turn now to the transformed Caldeira--Leggett's equation (\ref{rhoCL}),
\begin{align}\label{rhoCL}
 \frac{\partial}{\partial t}{\wti\rho}^{\text{\tiny HT}}_{\text{\tiny W}}
 & =-[i{\cal L}_{\tS}
      +({\wti L}^{\text{coh}}_{\FP}+{\wti L}^{\text{incoh}}_{\FP})]
        {\wti\rho}^{\text{\tiny HT}}_{\text{\tiny W}}
\nl&\quad
    +\Big(\frac{\alpha_1}{2}
    +\alpha_2x_{\B}\Big)
   \Big\{\hat Q_{\tS}, \wti{\frac{\partial}{\partial p_{\B}}}
    {\wti\rho}^{\text{\tiny HT}}_{\text{\tiny W}}
   \Big\}
 \nl & \quad
   -i\Big(\alpha_1x_{\B}
     +\alpha_2x_{\B}^2-\frac{\alpha_2}{4} \wti{\frac{\partial^2}{\partial p_{\B}^2}}\Big)
     [\hat Q_{\tS},
     {\wti\rho}^{\text{\tiny HT}}_{\text{\tiny W}}],
\end{align}
for
\begin{align}\label{wtirhoHT}
  {\wti\rho}^{\text{\tiny HT}}_{\text{\tiny W}}(x_{\B},p_{\B};t)
&\equiv \psi^{-1}_0(p_{\B}){\hat\rho}^{\text{\tiny HT}}_{\text{\tiny W}}(x_{\B},p_{\B};t)
\nl&
  = \sum_{n=0}^{\infty} \wti\rho_n(x_{\B};t)\psi_n(p_{\B}).
\end{align}
%%%
The second expression goes with
the complete and orthonormal basis set of $\{\psi_n(p_{\B}); n=0,1,\cdots\}$.
Therefore,
\be\label{wtirhon}
 \wti\rho_n(x_{\B};t)= \int^{\infty}_{-\infty}\!\!{\rm d}p_{\B}\,
   \psi_n(p_{\B}){\wti\rho}^{\text{\tiny HT}}_{\text{\tiny W}}(x_{\B},p_{\B};t).
\ee
Together with \Eqs{wtiL_FP_aa_dg}--(\ref{wti_partial_p2}),
we obtain
%%%%%%
\begin{align}\label{rhonx}
  \frac{\partial}{\partial t}\wti\rho_n
&=-(i{\cal L}_{\tS} + n\zeta)\wti\rho_n
   -i(\alpha_1x_{\B}+\alpha_2x_{\B}^2)[\hat Q_{\tS},\wti\rho_n]
\nl & \quad
     -\w_{\B} \sqrt{n+1}\sqrt{\beta\w_{\B}}(\hat D-x_{\B})\wti\rho_{n+1}
\nl & \quad
    -\w_{\B} \sqrt{n}\sqrt{\beta\w_{\B}}\hat D\wti\rho_{n-1}
\nl & \quad
    -\sqrt{n}\sqrt{\beta\w_{\B}}\Big(\frac{\alpha_1}{2}
    +\alpha_2x_{\B}\Big)
   \big\{\hat Q_{\tS}, \wti\rho_{n-1}\big\}
\nl & \quad
   +i \sqrt{n(n-1)}\,\frac{\alpha_2}{4} \beta\w_{\B}[\hat Q_{\tS},\wti\rho_{n-2}],
\end{align}
where
\be\label{app_hatD}
 \hat D\equiv (\beta\w_{\B})^{-1}
            \frac{\partial}{\partial x_{\B}}+x_{\B}.
\ee

 Consider now the Smoluchowski limit, where $\zeta\gg\w_{\B}$.
To derive a single closed equation in this  limit,
it requires further\cite{Ris89} (\emph{i})
 the core Hamiltonian, $H_{\tS}+\hat Q_{\tS}(\alpha_1x_{\B}+\alpha_2x_{\B}^2)$,
could be neglected compared to the friction, $\zeta$;
 (\emph{ii}) $\partial_t {\wti\rho}_{n>0}=0$; and (\emph{iii})
${\wti\rho}_{n>2}=0$, as the bath coupling is considered up to the quadratic level,
cf.\ \Eq{rhonx}.
Consequently, while
\begin{align}\label{app_rho1}
  \frac{\partial}{\partial t}\wti\rho_0&=
-i{\cal L}_{\tS}\wti\rho_0
   -i(\alpha_1x_{\B}
     +\alpha_2x_{\B}^2)
     [\hat Q_{\tS},
     \wti\rho_0]
\nl & \quad
     -\w_{\B}\sqrt{\beta\w_{\B}}(\hat D-x_{\B})\wti\rho_{1},
\end{align}
\Eq{rhonx} with $n=1$ and 2 becomes, respectively,
%%%%
\begin{align}\label{rho210}
0 &= -\zeta\wti\rho_1
     -\w_{\B} \sqrt{2}\sqrt{\beta\w_{\B}}(\hat D-x_{\B})\wti\rho_{2}
%\nl & \quad
    -\w_{\B}\sqrt{\beta\w_{\B}}\hat D\wti\rho_{0}
\nl & \quad
    -\sqrt{\beta\w_{\B}}\Big(\frac{\alpha_1}{2}
    +\alpha_2x_{\B}\Big)
   \big\{\hat Q_{\tS}, \wti\rho_{0}\big\},
%%%%
\\
  0&=
-2\zeta\wti\rho_2 -\w_{\B} \sqrt{2}\sqrt{\beta\w_{\B}}\hat D\wti\rho_{1}
\nl & \quad
    -\sqrt{2}\sqrt{\beta\w_{\B}}\Big(\frac{\alpha_1}{2}
    +\alpha_2x_{\B}\Big)
   \big\{\hat Q_{\tS}, \wti\rho_{1}\big\}
\nl & \quad
   +i \sqrt{2}  \,   \frac{\alpha_2}{4} \beta\w_{\B}
     [\hat Q_{\tS}, \wti\rho_{0}].
\end{align}
The latter two, together with \Eq{app_hatD} and $\gamma\equiv\w_{\B}^2/\zeta$, result in
\begin{align}\label{app_rho10}
 \zeta\wti\rho_1&=\frac{\w_{\B}}{\zeta}\frac{\partial}{\partial x_{\B}}
         \Big[\w_{\B}\hat D\wti\rho_1+
    \Big(\frac{\alpha_1}{2}
    +\alpha_2x_{\B}\Big)
   \big\{\hat Q_{\tS}, \wti\rho_{1}\big\}\Big]
\nl & \quad
     -i \frac{\alpha_2}{4} \frac{\beta\gamma}{\sqrt{\beta\w_{\B}}}\frac{\partial}{\partial x_{\B}}[\hat Q_{\tS}, \wti\rho_{0}]
%\nl & \quad
    -\w_{\B}\sqrt{\beta\w_{\B}}\hat D\wti\rho_{0}
\nl & \quad
    -\sqrt{\beta\w_{\B}}\Big(\frac{\alpha_1}{2}
    +\alpha_2x_{\B}\Big)
   \big\{\hat Q_{\tS}, \wti\rho_{0}\big\}
%%%%
\nl &\approx
     -i \frac{\alpha_2}{4} \frac{\beta\gamma}{\sqrt{\beta\w_{\B}}}\frac{\partial}{\partial x_{\B}}[\hat Q_{\tS}, \wti\rho_{0}]
%\nl & \quad
    -\w_{\B}\sqrt{\beta\w_{\B}}\hat D\wti\rho_{0}
\nl & \quad
    -\sqrt{\beta\w_{\B}}\Big(\frac{\alpha_1}{2}
    +\alpha_2x_{\B}\Big)
   \big\{\hat Q_{\tS}, \wti\rho_{0}\big\}.
\end{align}
The second expression is obtained by considering
 the Smoluchowski limit where $\zeta\gg\w_{\B}$.
Note also [cf.\ \Eqs{LD} and (\ref{app_hatD})]
\be\label{app_LD}
   \hat L_{\D}=-\gamma\frac{\partial}{\partial x_{\B}}\hat D,
\ee
and [cf.\ \Eq{eta_Zusman}]
\be\label{app_eta}
\begin{split}
   \eta_r&=\la x_{\B}^2\ra_{\B}=(\beta \w_{\B})^{-1},
\\
   \eta_i&=-\beta\gamma\la x_{\B}^2\ra_{\B}/2=-\gamma/(2\w_{\B}).
\end{split}
\ee
The high--temperature relation is used here; see comments after \Eq{calFP}.
Substituting \Eqs{app_rho10}--(\ref{app_eta}) into \Eq{app_rho1}, followed by
some simple algebra, we obtain
\begin{align}\label{app_exZE1}
  \frac{\partial}{\partial t}\wti\rho_{0}
=&- (i{\cal L}_{\tS}+{\hat L}_{\D})\wti\rho_{0}
%%%%%%
\nl&
   -i\Big[\alpha_1 x_{\B}+\alpha_2\Big(x_{\B}^2
   -\eta^{2}_i \frac{\partial^2}{\partial x_{\B}^2}\Big)\Big]
    \big[\hat Q_{\tS},\wti\rho_{0}\big]
\nl&
  - 2\eta_i \frac{\partial}{\partial x_{\B}}
    \left[\big(\frac{\alpha_1}{2}+\alpha_2 x_{\B}\big)
      \big\{\hat Q_{\tS},\wti\rho_{0}\big\}
    \right].
\end{align}
%%%%%%%%%%%%%%5
%%%%%%%%%%%%
\iffalse
\begin{align}
  \frac{\partial}{\partial t}\wti\rho_0&=
-i{\cal L}_{\tS}\wti\rho_0
   -i(\alpha_1x_{\B}
     +\alpha_2x_{\B}^2)
     [\hat Q_{\tS},
     \wti\rho_0]
%\nl & \quad
     +\gamma\frac{\partial}{\partial x_{\B}}\hat D\wti\rho_{0}
\nl & \quad
    +i\frac{\gamma}{\zeta}\frac{\alpha_2}{4}\frac{\partial^2}{\partial x_{\B}^2}[\hat Q_{\tS}, \wti\rho_{0}]
\nl & \quad
    +\frac{\w_{\B}}{\zeta}\frac{\partial}{\partial x_{\B}}\Big(\frac{\alpha_1}{2}
    +\alpha_2x_{\B}\Big)
   \big\{\hat Q_{\tS}, \wti\rho_{0}\big\}
,
\end{align}
\fi
%%%%%%%%%%%%%%%5
On the other hand, from \Eqs{rhoZusman}, (\ref{wtirhoHT}) and (\ref{wtirhon}), we have
\[
 \hat\rho(x_{\B};t)=\int_{-\infty}^\infty\!\! {\rm d} p_{\B}\,
    \psi_0(p_{\B}){\wti\rho}^{\text{\tiny HT}}_{\text{\tiny W}}(x_{\B},p_{\B};t)
 =\wti\rho_0(x_{\B};t).
\]
Therefore, \Eq{app_exZE1} is equivalent to \Eq{exZE1}.
We have thus completed the standard Fokker--Planck--Smoluchowski
approach\cite{Ris89} to the construction of the extended Zusman equation.
Apparently, the novel method of construction, on the basis of
\Eqs{Smolimit0}--(\ref{map0pm}) that result in the
rules of diffusion mapping, \Eq{highFriction_map}, is much simpler and physically more appealing.

\end{appendix}

%\bibliographystyle{../aiptit}
%\bibliography{../bibrefs}

\begin{thebibliography}{10}

\bibitem{Fey63118}
R.~P. Feynman and F.~L. \mbox{Vernon, Jr.}, \newblock ``The theory of a general
  quantum system interacting with a linear dissipative system,'' Ann. Phys.
  {\bf 24}, 118 (1963).

\bibitem{Tan906676}
Y.~Tanimura, \newblock ``Nonperturbative expansion method for a quantum system
  coupled to a harmonic-oscillator bath,'' Phys. Rev. A {\bf 41}, 6676 (1990).

\bibitem{Tan06082001}
Y.~Tanimura, \newblock ``Stochastic Liouville, Langevin, Fokker-Planck, and
  master equation approaches to quantum dissipative systems,'' J. Phys. Soc.
  Jpn. {\bf 75}, 082001 (2006).

\bibitem{Xu05041103}
R.~X. Xu, P.~Cui, X.~Q. Li, Y.~Mo, and Y.~J. Yan, \newblock ``Exact quantum
  master equation via the calculus on path integrals,'' J. Chem. Phys. {\bf
  122}, 041103 (2005).

\bibitem{Yan04216}
Y.~A. Yan, F.~Yang, Y.~Liu, and J.~S. Shao, \newblock ``Hierarchical approach
  based on stochastic decoupling to dissipative systems,'' Chem. Phys. Lett.
  {\bf 395}, 216 (2004).

\bibitem{Jin08234703}
J.~S. Jin, X.~Zheng, and Y.~J. Yan, \newblock ``Exact dynamics of dissipative
  electronic systems and quantum transport: Hierarchical equations of motion
  approach,'' J. Chem. Phys. {\bf 128}, 234703 (2008).

\bibitem{Wei08}
U.~Weiss,
\newblock {\em Quantum Dissipative Systems},
\newblock World Scientific, Singapore, 2008,
\newblock 3rd ed. Series in Modern Condensed Matter Physics, Vol.\ 13.

\bibitem{Kle09}
H.~Kleinert,
\newblock {\em Path Integrals in Quantum Mechanics, Statistics, Polymer
  Physics, and Financial Markets},
\newblock World Scientific, Singapore, 5th edition, 2009.

\bibitem{Yan05187}
Y.~J. Yan and R.~X. Xu, \newblock ``Quantum mechanics of dissipative systems,''
  Annu. Rev. Phys. Chem. {\bf 56}, 187 (2005).

\bibitem{Yan14054105}
Y.~J. Yan, \newblock ``Theory of open quantum systems with bath of electrons
  and phonons and spins: Many-dissipaton density matrixes approach,'' J. Chem.
  Phys. {\bf 140}, 054105 (2014).

\bibitem{Yan16110306}
Y.~J. Yan, J.~S. Jin, R.~X. Xu, and X.~Zheng, \newblock ``Dissipaton equation
  of motion approach to open quantum systems,'' Frontiers Phys. {\bf 11},
  110306 (2016).

\bibitem{Fan611866}
U.~Fano, \newblock ``Effects of configuration interaction on intensities and
  phase shifts,'' Phys. Rev. {\bf 124}, 1866 (1961).

\bibitem{Mir102257}
A.~E. Miroshnichenko, S.~Flach, and Y.~S. Kivshar, \newblock ``Fano resonances
  in nanoscale structures,'' Rev. Mod. Phys. {\bf 82}, 2257 (2010).

\bibitem{Zha15024112}
H.~D. Zhang, R.~X. Xu, X.~Zheng, and Y.~J. Yan, \newblock ``Nonperturbative
  spin-boson and spin-spin dynamics and nonlinear Fano interferences: A unified
  dissipaton theory based study,'' J. Chem. Phys. {\bf 142}, 024112 (2015).

\bibitem{Xu151816}
R.~X. Xu, H.~D. Zhang, X.~Zheng, and Y.~J. Yan, \newblock ``Dissipaton equation
  of motion for system-and-bath interference dynamics,'' Sci. China Chem. {\bf
  58}, 1816 (2015),
\newblock Special Issue: Lemin Li Festschrift.

\bibitem{Zha16237}
H.~D. Zhang, Q.~Qiao, R.~X. Xu, and Y.~J. Yan, \newblock ``Solvent-induced
  polarization dynamics and coherent two-dimensional spectroscopy: Dissipaton
  equation of motion approach,'' Chem. Phys. {\bf 481}, 237 (2016).

\bibitem{Zha16204109}
H.~D. Zhang, Q.~Qiao, R.~X. Xu, and Y.~J. Yan, \newblock ``Effects of
  Herzberg--Teller vibronic coupling on coherent excitation energy transfer,''
  J. Chem. Phys. {\bf 145}, 204109 (2016).

\bibitem{Jin15234108}
J.~S. Jin, S.~K. Wang, X.~Zheng, and Y.~J. Yan, \newblock ``Current noise
  spectra and mechanisms with dissipaton equation of motion theory,'' J. Chem.
  Phys. {\bf 142}, 234108 (2015).

\bibitem{Xu17JCP1}
R.~X. Xu, Y.~Liu, H.~D. Zhang, and Y.~J. Yan, \newblock ``Dissipaton equation
  of motion theory for a non--Gaussian bath coupling,'' J. Chem. Phys. {\bf
  xx}, xx (2017).

\bibitem{Cal83587}
A.~O. Caldeira and A.~J. Leggett, \newblock ``Path integral approach to quantum
  Brownian motion,'' Physica A {\bf 121}, 587 (1983).

\bibitem{Gar854491}
A.~Garg, J.~N. Onuchic, and V.~Ambegaokar, \newblock ``Effect of friction on
  electron transfer in biomolecules,'' J. Chem. Phys. {\bf 83}, 4491 (1985).

\bibitem{Ris89}
H.~Risken,
\newblock {\em The Fokker-Planck Equation, Methods of Solution and
  Applications},
\newblock Springer-Verlag, Berlin, 2nd edition, 1989.

\bibitem{Zha14319}
H.~D. Zhang, J.~Xu, R.~X. Xu, and Y.~J. Yan, \newblock ``Modified Zusman
  equation for quantum solvation dynamics and rate processes,'' in {\em
  Reaction Rate Constant Computations: Theories and Applications}, edited by
  K.-L. Han and T.-S. Chu, pages 319--336, Ch.\ 13, RSC Theoretical and
  Computational Chemistry Series No.6, London, 2014,
\newblock http://dx.doi.org/10.1039/9781849737753-00319.

\bibitem{Zus80295}
L.~D. Zusman, \newblock ``Outer-sphere electron transfer in polar solvents,''
  Chem. Phys. {\bf 49}, 295 (1980).

\bibitem{Zus8329}
L.~D. Zusman, \newblock ``The theory of transitions between electronic states.
  Application to radiationless transitions in polar solvents,'' Chem. Phys.
  {\bf 80}, 29 (1983).

\bibitem{Yan89281}
D.~Y. Yang and R.~I. Cukier, \newblock ``The transition from nonadiabatic to
  solvent controlled adiabatic electron transfer: Solvent dynamical effects in
  the inverted regime,'' J. Chem. Phys. {\bf 91}, 281 (1989).

\bibitem{Yan865908}
Y.~J. Yan and S.~Mukamel, \newblock ``Eigenstate-free, Green function:
  Calculation of molecular absorption and fluorescence line shapes,'' J. Chem.
  Phys. {\bf 85}, 5908 (1986).

\end{thebibliography}

\end{document}